\documentstyle[12pt]{article} 
\title{NEW EXPRESSION FOR THE TRANSVERSE DEFLECTION OF
RELATIVISTIC PARTICLE IN HIGH-FREQUENCY FIELDS AND CORRELATION WITH
PANOFSKY-WENZEL THEOREM}
\begin{document}
\author{V.N.Melekhin}
\date{15 January 1997}
\maketitle

{\bf Abstract}

It is shown that change in transverse momentum of a
relativistic particle, crossing an accelerating cavity parallel to
its axis ($z-$axis), may be presented as an integral over trajectory, 
the
integrand of which is proportional to $z-$component of magnetic 
field.  The
changes in $x-$ and $y-$components of momentum are equal in value but
opposite in sign. The obtained result is compared with
Panofsky-Wenzel theorem.

\vspace{5mm}
Some general theorem was demonstrated by Panofsky and Wenzel [1] for a
beam of fast particles passing through a cavity parallel to its axis.
The transverse momentum $p_{\perp}$ imparted to the particle (charge
$e$) can be presented as

$$
p_{\perp} =e \int_0^d \nabla_{\perp}(A_z) \cdot dz, \eqno (1)
$$

where $z$ is the distance along the axis of the cavity, the ends of
which corresponds to $z=0,\;d$ and ${\bf A}$ is the vector
potential. This equation was derived for the special case of a cavity
being some part of waveguide and having no transit holes. However, in
paper [2] eq.(1) was extended to the common case, when cavity has an
arbitrary shape and has transit holes.

For any TE-mode (no component of electric field ${\bf E}$ parallel to
the axis) eq.(1) gives $p_{\perp}=0\;$. Here and throughout this paper
the physical reason of such a result is as follows: the actions of 
transverse
electric and magnetic fields cancel each other being integrated over
trajectory. For a TM-mode (no component of magnetic field ${\bf H}$ 
parallel to
the axis) the equation (1) was presented in paper [1] as some integral
of $E_{\perp}$ only, independently of particle velocity.
However, this conclusion generally is in contradiction
with Bell's paper [3] in which it was shown that TM-mode, excited in 
circular
cavity with round transit holes, gives $p_{\perp} \to 0$ at $v_z \to 
c$. We
will discuss below the reason of such a discrepancy.

The latter result is of great importance for electron accelerators
because it inhibits, at first sight, high-frequency (HF) focusing of
electrons by accelerating field. However,
in our paper [4] the method of HF focusing was proposed which consists
in using not round but oval transit holes, may be slits, or
in using of noncircular cavities. In both cases the circular symmetry 
is
broken and in some direction the focusing by electric field exceeds
the magnetic field defocusing whereas in perpendicular direction one 
has the
reverse situation. This method was successfully used in
classical microtron (see [4] and [5]) and recently it was used for
calculation of race-track microtron [6].

In all these papers only
circular or rectangular cavities were considered. Let us calculate the
particle deflection in the common case with the only restriction that 
a
cavity has $x=0$ and $y=0$ planes of symmetry.  In this case, taking
the cavity symmetry into consideration, we can write approximate
formulae

$$
E_x=f_x x,\;\; E_y=f_y y,\;\; H_x=g_x y,\;\; H_y=g_y x,\;\;
H_z=g_z xy,  \eqno (2)
$$

where $x$ and $y$ are small deviations
of a particle trajectory from the cavity axis,
all coefficients $f$ and $g$ are certain functions of $z$ and
an electromagnetic field inside the cavity can be written as
${\bf E} cos(\omega t + \phi_{\circ})$,
${\bf H} sin(\omega t + \phi_{\circ})$ with a frequency $\omega$
and an arbitrary initial
phase $\phi_{\circ}$.  Now one can obtain the following relations,
using one of the Maxwell equations:

$$
g_z-\frac{dg_x}{dz} = k f_y,\;\; \frac{dg_y}{dz}-g_z=k f_x, \eqno (3)
$$

where $k=\frac{\omega}{c}$ is wave number.

Integrating by parts, we can write

$$
\int_0^d g_{\perp} \sin(\frac{kz}{\beta})\,dz =
\frac{\beta}{k} \left[g_{\perp}(0)-g_{\perp}(d)\cos(\frac{kz}{\beta})
+\int_0^d \frac{dg_{\perp}}{dz}\cos(\frac{kz}{\beta})\,dz \right ],
\eqno (4)
$$

where
$\frac{k z}{\beta}=\omega t + \phi_{\circ}$,
$\beta =\frac{v_z}{c} \approx\!const$ and $g_{\perp}$ is
$g_x$ or $g_y$. First and second terms in eq.(4) reduce to zero if
the initial $(z=0)$ and final $(z=d)$ points of trajectory are
situated outside the cavity.  Taking eqs. (3) and (4) into
account, one can derive:

$$
\frac{p_x}{x} = \frac{e}{\omega \beta}
\int_0^d \left [ (1-\beta^2) \frac{dg_y}{dz}-g_z \right ]
\cos(\frac{kz}{\beta})\,dz ,    \nonumber
$$

$$
\frac{p_y}{y} = -\frac{e}{\omega \beta}
\int_0^d \left [ (1-\beta^2) \frac{dg_x}{dz}-g_z \right ]
\cos(\frac{kz}{\beta})\,dz.  \eqno (5)
$$

We see that at $\beta \to 1$, that usually takes place in electron
accelerators, the first terms of the integrands in both equations
vanish, the result depends only on $z-$component of magnetic field and
we have the following relationships:

$$
\frac{p_x}{x} = - \frac{p_y}{y} =
-\frac{e}{\omega \beta}
\int_0^d g_z \cos(\frac{kz}{\beta})\,dz .    \eqno (6)
$$

The obtained result is interesting in some respects. First, one can 
see
that at any shape of cavity and transit holes
HF focusing in some direction
is accompanied by defocusing in transverse direction
the same as it takes place for usual quadrupole focusing.
Next, it follows from eqs.(6) that $p_{\perp}=0$
for TM-mode at $v_z \to c$, the same as it follows from eq.(1) for
TE-mode at any velocity. Such a
TM-mode may arise, for example, in a cavity of cylindrical symmetry 
and
this result corresponds to that of paper [3].  Such mode may also be
excited in rectangular cavity having such transit holes that 
parameters
$G=\alpha'=\alpha''$ which appear in eq.(4.13) of monograph [5].  It 
is
worth mentioning that first of the relations (6) follows from eqs.
(4.13) and (4.14) of this monograph for the calculated there circular
and rectangular cavities.

This result, concerning TM-mode, obviously is in contradiction with 
the
formula (7) of paper [1], in which $p_{\perp} \ne 0$ at any velocity.
Such a discrepancy arises not only due to different boundary 
conditions. The
mentioned formula (7) is incorrect because it was derived from the 
correct
equation (3) of paper [1], which corresponds to eq.(1) of this paper, 
under the
assumption $\nabla_{\perp}(A_z) \propto E_{\perp}$ (see eq.(4) in 
paper [1]),
which is valid only for a waveguide, not for a cavity. In any cavity 
mode the
mentioned values are displaced in time by quarter of a period. At the 
same time
eq.(6) is agreed upon the equation (1) first obtained by Panofsky and 
Wenzel
[1].

Here the equations (6) were derived for a cavity of an arbitrary
shape having two planes of symmetry. In this common case TM-modes also
can exist if the cavity is extended in some of these planes as much
that it corresponds to transit holes asymmetry. If there is no such an
accordance then magnetic lines penetrate into holes from the cavity 
and
week component $H_z$ arises. However, despite low level of this
component and the fact that Lorentz force does not depend on $H_z$, 
the
resulting focusing at $\beta \to 1$ is proportional to $H_z$ as it
follows from eq.(6). So, it follows from paper [1] together with this
paper, that high-energy electrons can be deflected or focused only
by such HF fields that have all six components and, hence, they are 
nor
TE- nor TM-modes.

The relations (6) are of practical importance for the
numerical calculation of HF focusing.
Such calculation is hampered by the fact that
great focusing and defocusing impulses, produced by quasi-static
electric field near transit holes and by HF magnetic field inside the
cavity, are subtracted from each other and final value is little
compared to initial ones. For this reason one needs to know the field
distribution with very high accuracy that is conjectural.  Taking this
into account, the equations (6) can be used to check a result
of numerical calculations.

I am grateful to V.I.Shvedunov and N.P.Sobenin called my attention to
the papers [1] and [2].

\end{document}